\documentstyle[referee]{mn}
\newcommand{\reference}{\bibitem}
\newcommand{\muas}{{\rm {\mu}as}}
\newcommand{\beq}{\begin{equation}}              
\newcommand{\beqa}{\begin{eqnarray}}             
\newcommand{\eeq}{\end{equation}}                
\newcommand{\eeqa}{\end{eqnarray}}               
\newcommand{\eeqi}[1]{\quad#1\end{equation}}     
\newcommand{\eeqai}[1]{\quad#1\end{eqnarray}}    

\newcommand{\ii}{\mbox{i}}
\newcommand{\dd}{\mbox{d}}
\newcommand{\sii}{\mbox{\scriptsize i}}

\input epsf

\title[]
{Extended Source Effects in Astrometric Gravitational Microlensing}
\author[Mao \& Witt]
{Shude Mao$^1$\thanks{e-mail: smao@mpa-garching.mpg.de}
Hans J. Witt$^2$\thanks{e-mail: hwitt@aip.de}  \\
$^1$Max-Planck-Institute f\"ur Astrophysik,
        Karl-Schwarzschild-Strasse 1, 85740 Garching, Germany \\
$^2$Astrophysikalisches Institut Potsdam, An der Sternwarte 16, 
        14482 Potsdam, Germany}

\date{Accepted ........
      Received .......;
      in original form .......}
\pagerange{\pageref{firstpage}--\pageref{lastpage}}
\pubyear{1998}

\begin{document}
\maketitle
\label{firstpage}
\begin{abstract}
Extended source size effects have been detected in photometric
monitoring of gravitational microlensing events.
We study similar effects in the centroid motion of
an extended source lensed by a point mass.
We show that the centroid motion of a source with
uniform surface brightness can be obtained analytically.
For a source with circularly symmetric 
limb-darkening profile, the centroid motion can be expressed as a
one-dimensional integral, which can be evaluated numerically.
We found that when the impact parameter is
comparable to the source radius, the centroid motion is significantly
modified by the finite source size.
In particular, when the impact parameter is
smaller than the source radius, the trajectories become clover-leaf
like. Such astrometric motions can be detected using
space interferometers such as the Space Interferometry Mission.
Such measurements offer exciting
possibilities of determining stellar parameters such as stellar radius
to excellent accuracy.
\end{abstract}


\section{Introduction}

Gravitational microlensing has produced many interesting scientific
returns concerning the Galactic structure, dark matter and variable stars
(see Paczy\'nski 1996a for a review). The current (photometric)
observations of microlensing events have one notable shortcoming: from
the observed time scale of microlensing event one cannot infer
the lens mass and
distance, since this measured quantity depends on the lens mass,
distances to the lens and the source and the relative lens and source
transverse motion.
This degeneracy can be partially or even completely 
removed using some rare
microlensing events (e.g., Gould 1996, 1997).
A much more general way of lifting the degeneracy is conducting an
accurate astrometric
program together with the photometric observations (
Hosokawa et al. 1993,
H${\rm \o}$g, Novikov \& Polnarev 1995, Miyamoto \& Yoshi 1995, and Walker 
1995); we refer to such an astrometric experiment as astrometric microlensing.
Accurate astrometry
can measure the angular Einstein radius and the relative parallax
between the lens and source. With these two additional
constraints, the lens mass can be determined, even if we have no
independent measures of the lens and source distances (Paczy\'nski 1998;
Boden, Shao, \& Van Buren 1998).
This method also opens up the possibility of measuring the mass of
nearby high proper-motion stars (Miralda-Escud\'e 1996,
Paczy\'nski 1996b). Boden et al. (1998) showed that 
the required astrometric accuracy can be reached by
ground interferometers soon available on 8-10m class telescopes such as
KECK and VLT. The Space Interferometry mission (SIM), to be launched in
2004, offers the best hope for
disentangling the lens degeneracy with its
global astrometric accuracy of $\sim \muas$.
Ground and space interferometers (for a review see Shao \& Colavita 1992)
therefore offers exciting possibilities of measuring the mass of lenses,
which may turn out to be brown dwarfs, planets etc.

Paczy\'nski (1998) pointed out that, using astrometric observations,
it is possible to measure the
stellar radius to 1\% accuracy for microlensed sources
which show finite size effect. The photometric extended source effect
has been studied by many workers
(Witt \& Mao 1994; Gould 1994; Nemiroff \& Wickramasinghe 1994).
However, up to now, most studies of astrometric microlensing
adopt a point source approximation; the only exception
is Walker (1995), who briefly discussed the astrometric extended source 
effect. Since the photometric extended source effects have already been
detected for several sources
(95-BLG-30, Alcock et al. 1997; 97-BLG-28, Albrow et al. 1998 and 
97-BLG-56) and in light of the possibility of measuring stellar radii
and other stellar parameters to
excellent accuracy, we study the astrometric extended source effects in
detail. In \S 2,
we first derive analytical expressions for the centroid
motion for a source with uniform surface brightness. The centroid motion
for sources with (realistic) limb-darkening profiles are then simplified
into a one-dimensional integral, which can be computed easily numerically.
In \S 3, we discuss briefly our results. Throughout this paper, we 
ignore parallactic effects, since this subject has been covered
in detail by other authors (e.g., Hosokawa et al. 1993,
H${\rm \o}$g et al. 1995, Miyamoto \& Yoshi 1995).

\section{Results}

We shall work in angular coordinates and all angles are normalized
to the Einstein ring angular scale
given by (e.g., Schneider, Ehlers, \&
Falco 1992; Paczy\'nski 1996a)
\begin{equation} \label{thetaE}
\theta _E = 
 0.9 ~ {\rm mas} ~ \left( { M \over M_{\odot} } \right) ^{1/2}
\left({ 10 ~ {\rm kpc} \over D } \right) ^{1/2} , 
{D \equiv {D_{d} D_s \over D_s-D_d}},
\end{equation}
where $ D_s $, $ D_d $, and $ M $ are the distance to the source and the
distance to 
the lens (deflector), and the lens mass, respectively. For convenience,
we first choose a coordinate system centered on the lens but later
(for observation considerations) transform the results
into a system where the source is at the origin
(see eq. [\ref{s-origin}]).
We use complex quantities to obtain a more
compact representation of the lens equation (Witt 1990). In this
formalism, the lens equation is simply
\beq \label{lens}
\zeta = z -\frac{1}{\bar{z}}
\eeqi{,}
where $z=x+\ii y$ are the coordinates in the deflector plane,
$\zeta =\xi +\ii \eta$ the coordinates in the source plane, and
$\bar{z}$ denotes the complex conjugate of $z$. Again,
both $z$ and $\zeta$ are normalized to $\theta_E$
given by Equation (\ref{thetaE}). (Note that for the case where
$z$ and $\zeta$ are expressed in length units we obtain a different 
normalization in each plane (cf. Witt \& Mao 1994)).

Eq. (\ref{lens}) can be easily solved to yield two image positions,
which are denoted by subscripts $+$ and $-$ indicating their respective
positive and negative parities (Witt \& Mao 1994):
\beq \label{images}
z_{+,-} = \frac{\zeta}{2} \left[1 \pm
\sqrt{1 +\frac{4}{\zeta\bar{\zeta}} }~ \right]
\eeqi{.}
The separation between the images are comparable to $\theta_E$ when
the distance between the lens and source is a few Einstein radii;
for microlensing in the local group, $D \sim 10\,{\rm kpc}$, the
separation is $\sim$ mas, too small to be resolved. Therefore we can only
observe the motion of the centroid of light for these two images. For a
point source, the centroid motion forms an ellipse in a coordinate sytem
where the unpreturbed source is at the origin (e.g., Walker
1995; see eq. [\ref{p-source}] below). In the
next subsections, we first study the centroid motion for 
extended sources with uniform
surface brightness and then for sources with more realistic profiles.

\subsection{Analytical Results for a Uniform Source}

Let us now consider a circular source with radius $r$ and constant
surface brightness. The boundary of the circular source is simply
described by
$\zeta(\varphi) =\zeta_0 + r e^{\sii\varphi}$.
Using eq. (\ref{images}),
we obtain a parametric representation for the image contours:
\beqa \label{circ}
z_{+,-} (\varphi ) &=& \frac{\zeta_0+r e^{\sii\varphi}}{2} \left[
 1\pm \sqrt{1 +\frac{4}{g(\varphi)}}
~ \right] \nonumber  \\
  &=& x_{+,-} (\varphi ) + \ii y_{+,-} (\varphi ),
\qquad 0 \le \varphi \leq 2\pi,
\eeqa
where we have assumed 
(without losing generality) that $\zeta_0$ is on the positive
real $\xi$-axis, 
and $g(\varphi) = r^2 +\zeta_0^2 +2r \zeta_0 \cos \varphi$.

Since lensing conserves surface brightness, the magnification of a circular
source with uniform surface brightness is just given 
by the ratio of the area of the images to the area of the source:
\beq \label{mag}
\mu_{\rm tot} = \frac{1}{\pi r^2} \int\limits_0^{2 \pi} \left[
-y_+(\varphi)\frac{\dd x_+(\varphi)}{\dd\varphi}
+y_-(\varphi)\frac{\dd x_-(\varphi)}{\dd\varphi}
\right]\dd \varphi
\eeqi{.}
Note that the minus sign results from the fact that 
when $\varphi$ changes from 0 to
$2\pi$, the contour for the image with
positive parity moves counter-clockwise whereas that with negative parity
moves clockwise.
This integral can be evaluated analytically; the results have been
presented in Witt \& Mao (1994). They are given below
for completeness and comparison purposes.

We can now calculate the centroid of light by weighting the
image positions with magnification. For a circular source we obtain
\beq \label{clight}
\Delta \theta_x = \frac{1}{\pi r^2 \mu_{\rm tot}} \int\limits_0^{2 \pi} 
\left[
- x_+(\varphi) y_+(\varphi)\frac{\dd x_+(\varphi)}{\dd\varphi}
+ x_-(\varphi) y_-(\varphi)\frac{\dd x_-(\varphi)}{\dd\varphi}
\right]\dd \varphi
\eeqi{.}
Obviously we have $\Delta\theta_y=0$ because of the reflection symmetry
with respect to the $x$-axis.

After a partial integration the integral becomes
\beq \label{partial}
\Delta \theta_x  = \frac{1}{3 \pi r \mu_{\rm tot}} \int\limits_0^{2 \pi} 
(\zeta_0+r \cos \varphi) (r+\zeta_0 \cos \varphi) 
\frac{1+g(\varphi)}{g(\varphi)}
\sqrt{1+\frac{4}{g(\varphi)} } \dd \varphi,
\eeq
where $g(\varphi)$ is defined under eq. (\ref{circ}).
After some tedious algebra we find, 
similar to the case of magnification for a uniform source,
the integral can be simplified into two expressions, one for the special
case of $\zeta_0 = r$ and one for the more general case when
$\zeta_0 \neq r$. 

For $\zeta_0 \neq r$, one finds
\beqa \label{neq}
\Delta \theta_x & = & \frac{1}{4 \pi r^2 \zeta_0 \mu_{\rm tot}
\sqrt{4+(\zeta_0-r)^2}} \left[ 
a_1 F(\frac{\pi}{2} ,k) + a_2 E(\frac{\pi}{2} ,k) +
a_3 \Pi (\frac{\pi}{2} ,n,k) \right], \\
\mu_{\rm tot} &= &
\frac{1}{2 \pi r^2 
\sqrt{4+(\zeta_0-r)^2}} \left[ 
b_1 F(\frac{\pi}{2} ,k) + b_2 E(\frac{\pi}{2} ,k) +
b_3 \Pi (\frac{\pi}{2} ,n,k) \right],
\eeqa
where $F, E, \Pi$ are respectively the first, second and third kind
of elliptic integrals as defined in Gradshteyn \& Ryzhik (1980), and 
\beq n= \frac{4\zeta_0 r}{(\zeta_0+r)^2}\quad\mbox{and} \quad
k= \sqrt{\frac{4 n}{4+(\zeta_0 -r)^2}}
\eeqi{.}  
The coefficients are defined as follows
\beqa
a_1 =& - (8+r^2+\zeta_0^2)(r+\zeta_0)(\zeta_0-r)^2,
 & b_1 = - (8-r^2+\zeta_0^2)(\zeta_0-r), \\
a_2 =& (4+(r-\zeta_0)^2)(r+\zeta_0)(r^2+\zeta_0^2),
 & b_2 =(4+(r-\zeta_0)^2)(r+\zeta_0), \\
a_3 =& 4(2r^2\zeta_0^2+r^2+\zeta_0^2)(r-\zeta_0)^2/(r+\zeta_0),
 & b_3 = 4(1+r^2)(r-\zeta_0)^2/(r+\zeta_0).
\eeqa
For $\zeta_0 = r$, we obtain
\beq \label{eq}
\Delta \theta_x = \zeta_0=r, ~~
\mu_{\rm tot} = \frac{2}{\pi}\left[
\frac{1}{r} + \frac{1+r^2}{r^2} \arctan r \right]
\eeqi{.}
The centroid motion for this special case is particularly
simple. Physically all images of the source boundary
fall on two circles: one circle is just the Einstein ring centered at the
origin and the other is centered on the source
with radius $(1+\zeta^2_0)^{1/2}$.
These two circles cross each other at two points, $(0, 1)$
and $(0, -1)$.
The contour for the positive parity images rotates counter-clockwise
along the two right arc segments and that for the negative party
images rotates clockwise along the two left arc segments. One can show
the resulting central shift of these simple configurations is just $\zeta_0$.

While it is possible to define the inertial system where
the (unperturbed) 
source is at rest, e.g., by using global astrometry or the positions of
nearby stars, it is often difficult or impossible to do so
for the lens since it could be dark. It is therefore observationally
convenient to express the centroid motion in a system where the
(unperturbed) source is
at the origin and the lens is moving, say,
along the $x$-axis. Each trajectory is
described by its impact parameter, $p$. The centroid position
can be obtained as
\beq \label{s-origin}
\Delta\theta^\prime_x = x - \Delta\theta_x \frac{x}{\zeta_0},~~~
\Delta\theta^\prime_y = p - \Delta\theta_x \frac{p}{\zeta_0}
\eeqi{,}
where the lens is at $(x,p)$ and $\zeta_0=(x^2+p^2)^{1/2}$. Notice 
it follows from eq. (\ref{eq}) that for $\zeta_0=r$,
the centroid shift in this coordinate system is zero. 
These results for extended sources
should be compared with those for a point source
(e.g., Walker 1995)
\beq \label{p-source}
\Delta\theta^\prime_x = -\frac{x}{\zeta^2_0+2},~~~
\Delta\theta^\prime_y = -\frac{p}{\zeta^2_0+2}
\eeqi{.}
When the lens moves from $-\infty$ to $+\infty$ parallel to the $x$-axis,
the centroid motion forms an ellipse with
the major axis along the $x$ direction and the minor axis along the $y$-axis.

The (topological)
difference between the centroid motion of a point source and that of
an extended source is approximately only a function of $r/p$.
In Fig. 1, we illustrate the difference with one
example for a source with $r=0.5$. The centroid motions for four
lens trajectories are shown with
$p=+3, +0.8, +0.5$, and $+0.1$, respectively.
When the impact parameter is much larger than the source radius
(top left), the trajectory of an extended source nearly
coincides with that for a point source, as expected intuitively.
When the impact parameter decreases to $\ga r$ (top right),
the trajectory begins to
deviate from the point source approximation, with the minor axis
becoming smaller. For the special case when $r=p$ (bottom left),
the centroid motion goes through the origin; this occurs when the edge of the 
source touches the lens (cf. eqs. [\ref{eq}] and [\ref{s-origin}]). 
As the impact
parameter is further reduced, the trajectory becomes clover-leaf like,
going through the origin twice (bottom right) since the lens hits the
edge of the source twice. 

\begin{figure}
\epsfysize=12cm
\centerline{\epsfbox{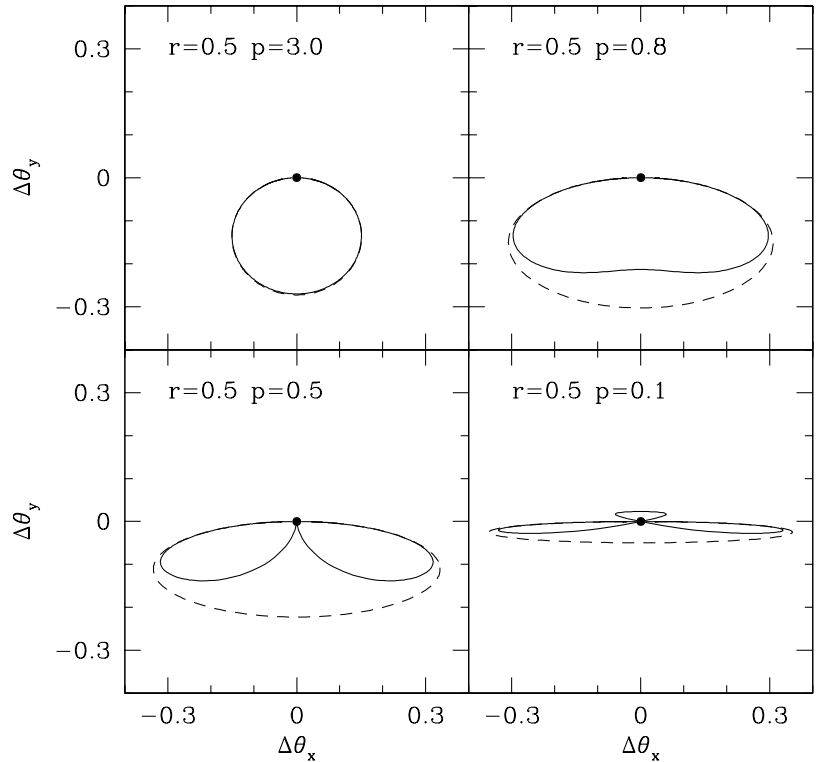}}
\caption{
Four examples of
astrometric trajectories showing finite source size effects. The source
size is assumed to be 0.5 Einstein radii ($r=0.5$).
For each example, the black dot marks the source position and
the lens is moving from $-\infty$ to $\infty$ parallel to the
$x$-axis. The centroid motion starts at the origin and moves
counter-clockwise. The impact
parameter (p) of the trajectory is shown at the left corner in each panel.
For each example, the solid line shows the trajectory that
takes into account the finite source size effect while the dashed lines shows
that for a point source approximation.
}
\end{figure}

\subsection{Generalization to Sources with Center-to-Limb Variations}

The surface brightness of stars in reality shows center-to-limb
variations. In this subsection, we extend our 
previous results (obtained for constant
surface brightness) to such cases. To do this, we first obtain
the magnification and centroid shift of an infinitely thin
shell at radius $R$. We illustrate the derivation by considering the 
magnification of a shell. 
The image area of a source with radius $R$
is given by $A(R) = \mu_{\rm tot}(R) \cdot \pi R^2$,
and that of a source with radius $R+dR$ 
is given by $A(R+dR) = \mu_{\rm tot}(R+dR) \cdot
\pi (R+dR)^2$. The shell
magnification is then just the differential image area divided by the
shell area, i.e., 
$[A(R+dR)-A(R)]/(2 \pi R dR)$, which yields
\beq
\mu_{\rm s} = \frac{1}{2\pi R} \frac{\dd}{\dd R}\left(\pi R^2 \mu_{\rm
tot}\right),
\eeq
where the subscript s indicates the quantities are for a thin shell.
Similarly for the centroid shift, we find
\beq
\Delta\theta_{x, {\rm s}}
=\frac{1}{2\pi R \mu_{\rm s}} \frac{\dd}{\dd R}\left(\pi R^2 \mu_{\rm tot}
\Delta\theta_x\right).
\eeq
Notice the extra magnification factors in $\Delta\theta_x$ arise because
the centroid is weighted by light.
For a source with circularly symmetric limb-darkening profile, $I(R)$, the
magnification and centroid shift are then simply
\beq
\mu_{\rm limb} =
\frac{\int\limits_0^{r} { \mu_{\rm s} I(R) 2\pi R \dd R}}
{\int\limits_0^{r} {I(R) 2 \pi R \dd R}}, ~~~
\Delta\theta_{x, {\rm limb}} = 
\frac{\int\limits_0^{r} {\Delta\theta_{x,{\rm s}} 
\mu_{\rm s} I(R) 2\pi R \dd R}}
{\int\limits_0^{r} {\mu_{\rm s} I(R) 2 \pi R \dd R}}
\eeqi{.}
These quantities can be easily evaluated for any limb-darkening profile
and transformed into other coordinate systems, e.g., using eq.
(\ref{s-origin}).

Fig. 2 shows the predicted centroid motion 
for the first microlensing event (95-BLG-30)
exhibiting (photometric) extended size effects discovered by the MACHO
collaboration (Alcock et al. 1997). 
The source limb-darkening profile is modelled by
\beq
\frac{I(R)}{I(0)}=
1-u_1-u_2 + u_1 \sqrt{1-\frac{R^2}{r^2}} + u_2 
\left(1-\frac{R^2}{r^2}\right),
\eeq
with $u_1=0.57, u_2=0.28$ for the MACHO V band, and 
$u_1=0.72, u_2=0.05$ for the MACHO R band.
The source radius and impact parameter are found to be $r =0.075,
p=0.055$, whereas the Einstein radius angular scale is
$\theta_E=420\muas$. For this event, the centroid motion displays the
clover-leaf like trajectory since we have $p< r$. The overall deviations can be
easily detected with SIM with its $\sim \muas$
astrometric accuracy. On the other hand, the centroid motions 
for sources with different profiles are almost the same.
We found that in almost all cases, the centroid
motion for a source with
limb-darkening profiles only show small differences from that for
a uniform source. Such small differences will be difficult to detect
astrometrically unless we have very dense time samplings.

\begin{figure}
\epsfysize=12cm
\centerline{\epsfbox{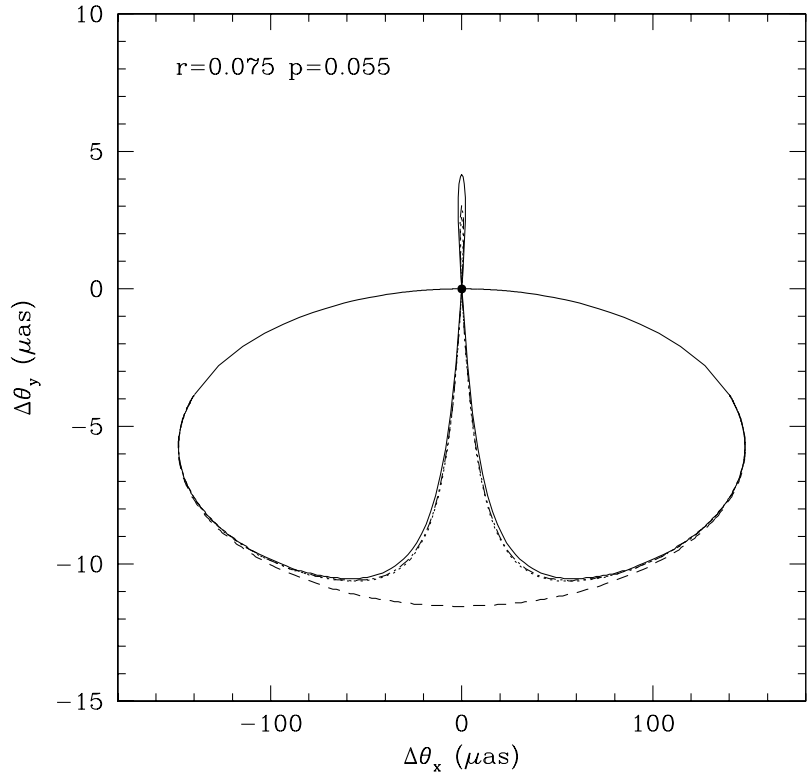}}
\caption{
Simulated astrometric trajectories for the first microlensing event
(95-BLG-30) that shows photometric extended source effects. The
parameters used are taken from Alcock et al. (1997).
The dashed ellipse is the centroid motion for a point source. 
The solid line shows the
trajectory for a source with constant surface brightness, whereas 
the dotted and dot-dashed lines show the predictions for the 
MACHO V and R band, respectively.
Note the scales are the two axes are {\it different} and should be
compared with the astrometric accuracy of $\sim \muas$ of the
Space Interferometry Mission.
}
\end{figure}

\section{Discussion} 

Extended source effects in photometric monitoring of
microlensing events have already been observed for several events. We
studied the astrometric signatures of these events in future
microlensing experiments using interferometry.
For microlensing events that exhibit extended source effects
photometrically, the impact
parameter is expected to 
be comparable to the source radius (e.g., Witt \& Mao 1994);
for such events, 
the centroid motion is significantly modified by the finite
source size effect and the distinctive clover-leaf
features can be easily detected with future interferometers such as SIM
(cf. Figs. 1 and 2). These modifications 
should be taken into account when fitting 
the centroid motions\footnote{The software for calculating the centroid
motion of an extended source is available at 
\it http://www.mpa-garching.mpg.de/\~\,smao/centroid}.
This will add one more parameter to the
fitting procedure, but this also means in principle the source size
can be obtained independent of
photometric observations. We conclude that using
astrometric microlensing, it is possible to determine interesting
stellar parameters such as stellar radius with $\sim 1\%$ accuracy
(Paczy\'nski 1998). We are anxiously waiting for the arrival of
astrometric microlensing.

\section*{Acknowledgments}
We thank Peter Schneider for a careful reading of the manuscript.
This work was partially supported by the ``Sonderforschungsbereich
375-95 f\"ur
Astro--Teil\-chen\-phy\-sik" der Deutschen For\-schungs\-ge\-mein\-schaft.


{}


\bsp
\label{lastpage}
\end{document}